 \documentclass[final,3p,times,preprint,review]{elsarticle}
\usepackage{comment}
\usepackage{caption}
\usepackage{subcaption}
\usepackage{tablefootnote}
\usepackage{amssymb}
\usepackage{amsmath}
\usepackage[boxed]{algorithm2e}
\usepackage{hyperref}
\hypersetup
{
    colorlinks=true,
    linkcolor=blue,
    filecolor=magenta,      
    urlcolor=cyan,
}
\journal{Energy and Buildings}

\begin{document}

\begin{frontmatter}



\title{FastCTF: A Robust Solver for Conduction Transfer Function Coefficients and Thermal Response Factors}


\author[inst1]{Khodr Jaber\corref{email1}}
\cortext[email1]{Corresponding author. Tel.: +974 3362 8564 \ead{jaberjbr2@gmail.com}}

% \ead[inst1]{jaberjbr2@gmail.com}
\affiliation[inst1]{organization={Meinhardt Group},%Department and Organization
            addressline={Amwal Tower}, 
            city={Doha},
            %postcode={00000}, 
            %state={State One},
            country={Qatar}}

%\author[inst2]{Author Two}
%\author[inst1,inst2]{Author Three}

%\affiliation[inst2]{organization={Department Two},%Department and Organization
%           {addressline={Address Two}, 
%            city={City Two},
%            postcode={22222}, 
%            state={State Two},
%            country={Country Two}}

\begin{abstract}
Conduction transfer functions (CTF) are commonly used in the building services to quickly estimate hourly conduction heat loads through multilayered walls without resorting to expensive, time-consuming solutions of the heat equation. It is essential for any software developed for this purpose to be able to simulate walls of varying weight with a high degree of accuracy. A robust algorithm for computing CTF coefficients and thermal response factors based on power series expansions of solutions of the governing equations in the complex $s$-domain is presented and validated. These series expansions are used to to construct Padé approximants of the system's transfer functions, which greatly simplifies the inversion of the solution from the complex domain to the time domain, and allows for an easy recovery of a time series representation via the Z-transform. The algorithm is also implemented in an open-source C++ code. Its performance is validated with respect to exact theoretical frequency characteristics and its results are compared with data generated by previously established methods for computing CTF coefficients / response factors.
\end{abstract}


\begin{highlights}
\item Open source solver for CTF coefficients and response factors
\item Series expansion estimation of hyperbolic conduction transfer function
\item Performance under various time steps and wall weights
\item Comparison with ASHRAE Handbook and Frequency-Domain Regression
\end{highlights}


\end{frontmatter}


\section{Introduction}
\label{sec:intro}

The need for efficient heat load calculation strategies in the building services has led to the development of a variety of numerical tools that allow for automated implementations which had previously been performed manually by hand. Examples of such strategies include the heat balance method \cite{HB} and its derivatives - the transfer function method \cite{TFM}, cooling load temperature difference and the radiant time series method \cite{RTS}. The heat balance method directly applies first principles and involves setting up an energy balance over conditioned spaces in terms of surface and interior temperatures / heat fluxes, resulting in a system of equations whose solution can be cumbersome due to transient interactions between internal surface conditions and element conduction effects driven by the heat equation (which is assumed to be 1D for roofs and walls):
\begin{align}
    \begin{cases}
        \displaystyle\frac{\partial T_k}{\partial t} &= \displaystyle\frac{\lambda_k}{\rho_k C_{p,k}}\frac{\partial^2 T_k}{\partial x^2} \\
        q_k &= \displaystyle -\lambda_k \frac{\partial T_k}{\partial x}
    \end{cases}, \quad k = 1,...,N_l \label{eq:gov-1}
\end{align}
\begin{align}
    \begin{cases}
        T_k(0,t) &= T_{k-1}(L_k,t) \\
        q_k(0,t) &= q_{k-1}(L_k,t)
    \end{cases}, \quad k = 2,...,N_l \label{eq:gov-2}
\end{align}

\noindent where $\lambda_k, \rho_k, C_{p,k}, T_k$ and $q_k$ denote thermal conductivity, density, specific heat capacity, temperature distribution and heat flux through the $k^{\text{th}}$ layer of an $N_l$-layer element. Direct simulation of the governing equations is made computationally infeasible by the stability constraints imposed on the choice of time step (which would need to be on the order of an hour to estimate yearly energy profiles in a reasonable amount of time). This constraint has motivated the development of methods based on time series representations of the conduction process which simplify the internal surface-conduction relationship and provide the basis for a time marching algorithm with a more feasible time step. The most popular of these methods are the \textit{response factor method} and \textit{transfer function method}, the latter which continues to be employed in well-known commercial software such as Carrier HAP as it does not require surface / interior conditions as input and does not place any constraint on outdoor conditions (as opposed to the RTS, which exploits the steady periodic nature of a day-by-day calculation process). The reader is referred to \cite{FDR} and \cite{Review} for an exhaustive historical review of transient heat flow calculation methods.

The response factor method is an approach in which heat fluxes are written as a time series in terms of temperature history:
\begin{align}
    q_{\text{ext.}}(n \Delta t) = \sum_{k=0}^\infty X[k] T_{\text{ext.}} [(n-k)\Delta t] - \sum_{k=0}^\infty Y[k] T_{\text{int.}} [(n-k)\Delta t] \\
    q_{\text{int.}}(n \Delta t) = \sum_{k=0}^\infty Y[k] T_{\text{ext.}} [(n-k)\Delta t] - \sum_{k=0}^\infty Z[k] T_{\text{int.}} [(n-k)\Delta t]
\end{align}

\noindent where $T_{\text{ext.}}, T_{\text{int.}}$ are external and internal temperatures of the element, and the coefficients $\{X\}_k, \{Y\}_k, \{Z\}_k$ are referred to as external, cross and internal response factors. These coefficients represent the response of a thermal element when subjected to temperature pulses in discrete time $\{ k \Delta t \}_k$. Brisken and Reque are credited \cite{FDR} with establishing the method \cite{Brisken}, which was later improved significantly by Mitalas and Stephenson \cite{Mitalas I, Mitalas II, Mitalas III}. It is noted that this time series could become infeasibly long, motivating Mitalas and Stephenson to develop the transfer function method \cite{TFM}, which allowed heat fluxes to be represented by a substantially shorter series in both temperature and flux  history:
\begin{align}
    q_{\text{net}} = \sum_k b_k T_{o,k-n} - \sum_k d_k q_{k-n} - T_{\text{rc}}\sum_k c_k
\end{align}

\noindent where $T_{o,k-n}$ is the outdoor temperature at $t = (k-n)\Delta t$, $T_{\text{rc}}$ is the required room temperature (assumed to b constant) and $b_k, c_k, d_k$ are the transfer function coefficients. To do this, Laplace transforms are usually applied to the governing conduction equations so that the analysis can be performed in the complex $s$-domain (although this need not be the case, as in Davies' time-domain analysis \cite{Davies}, or the state-space methodology which employs finite differences/element analysis to recover the conduction transfer function). Z-transforms are then used to transform the time-domain solution (which is obtained after applying an inverse Laplace transform) into a discrete-time transfer function (i.e. the z-transfer function). Direct root finding (DRF) was first developed to obtain the inverted solution by resolving the integral (inverse) transform via the residue theorem. DRF is used in simulation programs such as PREP (in TRNSYS \cite{TRNSYS}) and BLAST \cite{BLAST}, while the state space method is used in the EnergyPlus software \cite{EnergyPlus}. The RTS method of Spitler et al. is a special case of the response factor method and assumes that outside temperature can be put in the form of a 24-hour periodic set, so that the response factors can be written as:
\begin{align}
    q_{\text{net}}(n \Delta t) = \sum_{k=0}^{23} Y_{\text{RTS}}[k] \cdot T_{o, k-n} - T_{\text{rc}} \sum_{k=0}^{23} Z_{\text{RTS}}[k]
\end{align}

\noindent where $\{Y_{\text{RTS}}\}, \{Z_{\text{RTS}}\}$ are response factors for periodic conditions. It has been shown by Spitler and Fisher \cite{RTS-TFM} that these response factors can be derived from conduction transfer coefficients using linear algebra. More recently developed methods include Frequency-Domain Regression (FDR) by Wang and Chen \cite{FDR, FDR II} (which uses least squares regression in the frequency domain to simplify Laplace inversion), Direct Numerical Integration (DNI) by Varela et al \cite{DNI} (which provides an alternative numerical Laplace inversion strategy) and Frequency-Domain Spline Interpolation (FDSI) by Pérez et al. \cite{FDSI} (which employs Fourier analysis to recover the heat flux in the frequency domain).

\textit{FastCTF} is an open source C++ code capable of computing CTF coefficients and response factors. It implements a novel algorithm motivated by the FDR method, as presented in \cite{FDR mI, FDR}, in that the transfer functions of the system in the $s$-domain (consisting of a complex arrangement of hyperbolic trigonometric functions) are approximated in the form of rational functions:
\begin{align}
    G(s) \approx \frac{\sum_{k = 0}^m \widetilde{\beta}_k s^k}{\sum_{k = 0}^m \widetilde{\alpha}_k s^k}
\end{align}

This is done by approximating the $s$-domain solution via high-order Taylor series expansions and using them to construct rational functions representations of the external, cross and internal flow transfer functions via suitable order Padé approximations. Due to heavy use of series expansions in the s-plane, the algorithm will be referred to as the Complex Domain Series Expansion (CDSE) method for the remainder of this paper. Once Padé approximants of the transfer functions have been determined, the inverse Laplace transform is easily applied to recover the solutions in the time domain. Finally, Z-transforms are used to derive the z-transfer functions (from which the CTF coefficients are obtained). An appropriate partial fraction decomposition is performed via the residue method to facilitate the inverse Laplace transform. Forms of the time-domain solutions can be predicted in advance through knowledge of the poles and residues, so the z-transfer functions are assembled immediately after these have been calculated. 

The CDSE method is a purely mathematical exercise requiring only basic linear algebra for computing Padé approximants and polynomial roots. Iterative methods are not used at any point, eliminating potential numerical stability concerns. The conditioning of the relevant linear systems does not require any special care, even when approximants of up to order 10 are sought (although order 6 is sufficient in practice). Quality control of the results corresponds to tolerances on errors between the frequency characteristics of the CTF z-transfer functions and their theoretical counterparts. Analysis of the latter is formalized using an error criterion defined by Chen et al. \cite{FDR-Ver II} in their investigation of a verification and validation strategy for CTF coefficients and response factors.

The s-transfer function estimation procedure will be described first, followed by an illustration of the time-domain solution corresponding to a ramp input and its corresponding Z-transform. Steps for assembling the z-transfer functions are then shown, and it is demonstrated that the thermal response factors can be accurately recovered from them. Finally, a set of case studies will be investigated to verify the results of the CDSE formulation with respect to theoretical values and in comparison to other results reported in the literature (specifically, those generated by the FDR method and reported in ASHRAE's well-known Handbook for HVAC calculations).

\section{Modeling of Conduction Through a Multilayered Element}

Assuming that the physical properties of all layers in a conductive element are constant, the Laplace transform can be applied to governing equations (\ref{eq:gov-1}), (\ref{eq:gov-2}) to obtain a complex-domain solution of the form:
\begin{align}
    \begin{bmatrix}
        T_{\text{in}} \\
        q_{\text{in}}
    \end{bmatrix} &= \underbrace{\begin{bmatrix}
        A(s) & B(s) \\
        C(s) & D(s)
    \end{bmatrix}}_{\textbf{M}(s)} \begin{bmatrix}
        T_{\text{out}} \\
        q_{\text{out}}
    \end{bmatrix}
\end{align}

\noindent where the overall transmission matrix $\textbf{M}(s)$ can be written as a product of individual transmission matrices for each layer of the conductive element:
\begin{align}
    \textbf{M}(s) &= \textbf{M}_{\text{in}} \bigg( \prod_{i=1}^{N_l} \textbf{M}_j(s) \bigg) \textbf{M}_{\text{out}}
\end{align}
\begin{align}
    \textbf{M}_{\text{in}} = \begin{bmatrix}
        1 & R_{\text{in}} \\
        0 & 1
    \end{bmatrix}, \quad \underset{1\leq j\leq N_l}{\textbf{M}_j(s)} = \begin{bmatrix}
        \displaystyle \cosh \bigg( L_j \sqrt{\frac{s}{\alpha_j}} \bigg) & \displaystyle\sinh \bigg( L_j \sqrt{\frac{s}{\alpha_j}} \bigg) \bigg/ \bigg(\lambda_j \sqrt{\frac{s}{\alpha_j}}\bigg)\\[12pt]
        \displaystyle \bigg( \lambda_j \sqrt{\frac{s}{\alpha_j}} \bigg) \sinh \bigg( L_j \sqrt{\frac{s}{\alpha_j}} \bigg) & \displaystyle \cosh \bigg( L_j \sqrt{\frac{s}{\alpha_j}} \bigg)
    \end{bmatrix}, \quad \textbf{M}_{\text{out}}(s) = \begin{bmatrix}
        1 & R_{\text{out}} \\
        0 & 1
    \end{bmatrix}
\end{align}

\noindent and $\textbf{M}_{\text{in}}, \textbf{M}_{\text{out}}$ correspond to transmission matrices for the inner and outer surfaces films of the element, respectively. Transfer functions for the heat flux entering / leaving an element can be found by rewriting the system above as a function of the temperatures:
\begin{align}
    \begin{bmatrix}
        q_{\text{int.}} \\
        q_{\text{ext.}}
    \end{bmatrix} &= \begin{bmatrix}
        \displaystyle\frac{D(s)}{B(s)} && -\displaystyle\frac{1}{B(s)} \\[12pt]
        \displaystyle\frac{1}{B(s)} && -\displaystyle\frac{A(s)}{B(s)}
    \end{bmatrix} \begin{bmatrix}
        T_{\text{ext.}} \\
        T_{\text{int.}}
    \end{bmatrix} = \begin{bmatrix}
        G_X(s) & -G_Y(s) \\
        G_Y(s) & -G_Z(s)
    \end{bmatrix} \begin{bmatrix}
        T_{\text{ext.}} \\
        T_{\text{int.}}
    \end{bmatrix}
\end{align}

\noindent Here, $G_X(s), G_Y(s)$ and $G_Z(s)$ are the external, cross and internal transfer functions, respectively. Explicitly, we have:
\begin{align}
    G_X(s) = \frac{A(s)}{B(s)}, \ G_Y(s) = \frac{1}{B(s)}, \ G_Z(s) = \frac{D(s)}{B(s)}
\end{align}

\section{Numerical Methodology}

    \subsection{Series Expansions Formulation for the $s$-Transfer Function}
    
    The present aim is to construct rational function approximations of the s-transfer functions  of the form:
    \begin{align}
        G_{\phi}(s) = \frac{\sum_{k = 0}^m \widetilde{\beta}_{\phi,k} s^k}{1 + \sum_{k = 1}^m \widetilde{\alpha}_{\phi,k} s^k}, \quad \phi \in \{X, Y, Z\}
    \end{align}
    
    \noindent In doing so, computing the inverse Laplace transform becomes straightforward after performing partial fraction decomposition. To obtain this rational function approximation, Taylor series expansions are sought for elements of the individual transmission matrices. For $(M_j)_{1,1}$ and $(M_j)_{2,2}$, we have:
    \begin{align}
         \displaystyle\cosh{ \bigg( L^{[j]} \sqrt{s / \alpha^{[j]}} \bigg) } &= \sum_{k=0}^N \frac{\bigg(L^{[j]} \sqrt{s / \alpha^{[j]}}\bigg)^{2k}}{(2k)!} + \mathcal{O}\big(s^{N+1}\big)
    \end{align}
    
    \noindent where $L^{[j]}$ and $\alpha^{[j]}$ are the length and thermal diffusivity of the $j^{\text{th}}$ wall layer. To get the Taylor series expansions of the other matrix entries, a Puiseux series (that is, a power series with fractional exponents in the indeterminates) is sought for the hyperbolic sine function in terms of $\sqrt{s}$:
    \begin{align}
         \displaystyle\sinh{ \bigg( L^{[j]} \sqrt{s / \alpha^{[j]}} \bigg) } &= \sum_{k=0}^N \frac{\bigg(L^{[j]} \sqrt{s / \alpha^{[j]}}\bigg)^{2k+1}}{(1+2k)!} + \mathcal{O}\big(s^{N+1/2}\big)
    \end{align}
    
    \noindent In other words, the Taylor series of $\sinh{( L^{[j]} y) }$ is found, with $y$ substituted for $\sqrt{s / \alpha^{[j]}}$. Now, the square-root term can be divided/multiplied into this series to obtain $(M_j)_{1,2}$ and $(M_j)_{2,1}$, respectively:
    \begin{align}
         \displaystyle\sinh{ \bigg( L^{[j]} \sqrt{s / \alpha^{[j]}} \bigg) } \ \bigg / \ \bigg(\sqrt{s / \alpha^{[j]}}\bigg) &= \sum_{k=0}^{N} \frac{\bigg(L^{[j]}\bigg)^{2k+1} \bigg(\sqrt{s / \alpha^{[j]}}\bigg)^{2k}}{(1+2k)!} + \mathcal{O}\big(s^{N+1}\big) \\
         \bigg(\sqrt{s / \alpha^{[j]}}\bigg) \displaystyle\sinh{ \bigg( L^{[j]} \sqrt{s / \alpha^{[j]}} \bigg) } &= \sum_{k=0}^{N-1} \frac{ \bigg(L^{[j]}\bigg)^{2k+1} \bigg( \sqrt{s / \alpha^{[j]}}\bigg)^{2(k+1)}}{(1+2k)!} + \mathcal{O}\big(s^{N+1}\big)
    \end{align}
    
    Multiplication of the individual transmission matrices can now be performed in sequence:
    \begin{align}
         \textbf{M}_j &= \underbrace{\begin{bmatrix}
             \displaystyle \sum_{k=0}^{N} M_{11,k}^{[j]} s^k & \displaystyle \sum_{k=0}^{N} M_{12,k}^{[j]} s^k \\
             \displaystyle \sum_{k=0}^{N} M_{21,k}^{[j]} s^k & \displaystyle \sum_{k=0}^{N} M_{22,k}^{[j]} s^k 
         \end{bmatrix}}_{\textbf{M}^{[j]}} \textbf{M}_{j-1}, \quad j = 1,...,N_l+1 \\
         \textbf{M}_0 &= \textbf{M}_{\text{out}}, \quad \textbf{M}^{[N_l+1]} = \textbf{M}_{\text{in}}, \quad \textbf{M}_{N_l} = \textbf{M} \\
         \textbf{M} &= \begin{bmatrix}
             \displaystyle \sum_{k=0}^{N} M_{11,k} s^k & \displaystyle \sum_{k=0}^{N} M_{12,k} s^k \\
             \displaystyle \sum_{k=0}^{N} M_{21,k} s^k & \displaystyle \sum_{k=0}^{N} M_{22,k} s^k 
         \end{bmatrix}
    \end{align}
    
    To ensure that the resulting approximation of the Laplace-domain solution is of high accuracy, terms of up to order 20 are included in the Taylor series expansions. The external, cross and internal heat flow transfer functions are thus represented as ratios of these series entries, however, the numerical conditioning of a polynomial root-finding problem applied to the denominators of these rational functions at this stage would be poor and the method would become redundant. On the other hand, the unique structure of the cross flow transfer function can be exploited by realizing that it is just the multiplicative inverse of a Taylor series. With this series, an appropriate-order s-transfer function can be estimated via its corresponding $[n,m]$-order Padé approximant, defined by:
    \begin{align}
        \displaystyle\frac{1}{G_{Y}(s)} = \sum_{k=0}^{\infty} T_k s^k &\approx \frac{\sum_{k=0}^n P_k x^k}{\sum_{k=0}^m Q_k x^k}
    \end{align}

    \noindent which is computed by moving the left-hand-side into the fraction:
    \begin{align}
        \sum_{k=0}^{\infty} T_k s^k - \frac{\sum_{k=0}^n P_k x^k}{\sum_{k=0}^m Q_k x^k} &\approx \frac{\sum_{k=0}^{\infty} T_k s^k \sum_{k=0}^m Q_k x^k - \sum_{k=0}^n P_k x^k}{\sum_{k=0}^m Q_k x^k} = 0
    \end{align}
    
    \noindent and requiring that the coefficients of the numerator vanish ($Q_0 = 1$ is fixed to ensure uniqueness of the approximant):
    \begin{align}
        \begin{bmatrix}
            0 & 0 & ... & 0 & -1 & 0 & ... & 0 \\
            T_0 & 0 & ... & 0 & 0 & -1 & ... & 0 \\
            \vdots & & & & & & & \\
            T_{m-1} & T_{m-1} & ... & T_{m-n} & 0 & 0 & ... & -1 \\
            T_{m+1} & T_{m} & ... & T_{m-n+1} & 0 & 0 & ... & 0 \\
            T_{m+2} & T_{m+1} & ... & T_{m-n+2} & 0 & 0 & ... & 0 \\
            \vdots & & & & & & & \\
            T_{m+n} & T_{m+n-1} & ... & T_{m} & 0 & 0 & ... & 0 \\
        \end{bmatrix} \begin{bmatrix}
            Q_1 \\
            Q_2 \\
            \vdots \\
            Q_n \\[8pt]
            P_0 \\
            P_1 \\
            \vdots \\
            P_m
        \end{bmatrix} = \begin{bmatrix}
            -T_0 \\
            -T_1 \\
            \vdots \\
            -G_m \\[8pt]
            0 \\
            0 \\
            \vdots \\
            0
        \end{bmatrix}
    \end{align}
    
    Applying the multiplicative inverse is straightforward, and we now have:
    \begin{align}
        G_{Y}(s) \approx \frac{\sum_{k=0}^n Q_k x^k}{\sum_{k=0}^m P_k x^k}
    \end{align}
    
    \noindent It will be shown later that the z-transfer function that is computed from this approximant will be always have an $m^{\text{th}}$-order numerator and denominator, so the choice of $n \leq m$ is arbitrary ($n = m$ is fixed in the implementation and in the remainder of the method description for convenience).
    
    The Padé approximants for $G_X(s)$ and $G_Z(s)$ are derived using the polynomial $\sum_k P_k$ by equating their transfer functions with corresponding transmission matrix entries in the following way:
    \begin{align}
        G_X(s) &\approx \frac{\sum_{k=0}^N M_{11,k} x^k}{\sum_{k=0}^N M_{12,k} x^k} = \frac{\sum_{k=0}^m (Q_X)_k x^k}{\sum_{k=0}^m P_k x^k} \\
        G_Z(s) &\approx \frac{\sum_{k=0}^N M_{22,k} x^k}{\sum_{k=0}^N M_{12,k} x^k} = \frac{\sum_{k=0}^m (Q_Z)_k x^k}{\sum_{k=0}^m P_k x^k}
    \end{align}
    
    \noindent After clearing denominators on both sides of the equations, recurrence relations can be established for the $(Q_X)_k$ and $(Q_Z)_k$:
    \begin{align}
        (Q_X)_k &= \frac{1}{M_{12,0}} \bigg[ \bigg(\sum_{j=0}^N M_{11,j} x^j \sum_{j=0}^m P_j x^j \bigg)_k - \sum_{j=0}^{k-1} (Q_X)_j M_{12,k-j} \bigg], \quad k = 0,...,m \\
        (Q_Z)_k &= \frac{1}{M_{12,0}} \bigg[ \bigg(\sum_{j=0}^N M_{22,j} x^j \sum_{j=0}^m P_j x^j \bigg)_k - \sum_{j=0}^{k-1} (Q_Z)_j M_{12,k-j} \bigg], \quad k = 0,...,m
    \end{align}
    
    \subsection{Response Factors and the Conduction Transfer Function}
    
    The remainder of the calculation procedure is similar section 4 of \cite{FDR} albeit with a  modified Laplace inversion formula that accounts for complex poles in the transfer functions (real roots were guaranteed in the FDR method since the rational function approximation was fitted to the frequency response of the system). The generalized Laplace inversion formula for a system subjected to a ramp input can be written as:
    \begin{align}
         \frac{\sum_{k=0}^r \widetilde{\beta}_k s^k}{s^2(1+\sum_{k=1}^m \widetilde{\alpha}_k s^k)} &= \frac{K_1}{s^2} + \frac{K_2}{s} + \sum_{k=1}^{\Lambda_R} \frac{\gamma_k}{s + r_k} + \sum_{k=1}^{\Lambda_I} \bigg( \frac{\varsigma_1 + j \varsigma_2}{s + \sigma_{1,k} + \sigma_{2,k} j} + \frac{\varsigma_1 - j \varsigma_2}{s + \sigma_{1,k} - \sigma_{2,k} j} \bigg) \\
         q(t) = \mathcal{L}^{-1}\bigg[ \frac{\sum_{k=0}^r \widetilde{\beta}_k s^k}{s^2(1+\sum_{k=1}^m \widetilde{\alpha}_k s^k)} \bigg] &= K_1 t + K_2 + \sum_{k=1}^{\Lambda_R} \gamma_k e^{-r_k t} + \sum_{k=1}^{\Lambda_I} 2e^{-\sigma_{1,k} t}\big( \varsigma_1 \cos(\sigma_{2,k} t) + \varsigma_2 \sin(\sigma_{2,k} t) \big)
    \end{align}
    
    \noindent where $r_k$ are poles of the $s$-transfer function (with $r_k = \sigma_1 + j \sigma_2$ for complex cases) and $\gamma_k, \varsigma_{1/2, k}$ denote residues in the partial fraction decomposition corresponding to $\Lambda_R$ real roots and $\Lambda_I$  pairs of complex roots (that is, a complex root and its conjugate are a single element counted by $\Lambda_I$ such that $\Lambda_R + 2\Lambda_I = m$ ), respectively.
    
    This ramp response is used to construct  triangular pulse responses that define the response factors. Three ramps at $t-\Delta t, t, t + \Delta t$ with slopes of $\Delta t^{-1}, -2\Delta t^{-1}, \Delta t^{-1}$ are superimposed, and the response of the system at time $t = k \Delta t$ defines the $k^{\text{th}}$ response factor $Y[k]$:
    \begin{align}
        \theta(t) &= \frac{1}{\Delta t}\big[q(t - \Delta t) - 2q(t) + q(t + \Delta t)\big] \label{eq:theta_1} \\
        Y[0] &= \frac{1}{\Delta t}q(\Delta t) \label{eq:theta_2} \\
        Y[k] &= \frac{1}{\Delta t}[q((k-1)\Delta t) - 2q(k\Delta t) + q((k+1)\Delta t)], \quad k = 1,2,... \label{eq:theta_3}
    \end{align}
    
    \noindent The function $\theta(t)$ is the heat flow due to a triangular pulse excitation in the continuous time domain. Applying the Z-transform to this function recovers the transfer function in the z-domain:
    \begin{align}
        G^Z_{\phi}(z) = Z\big[\theta\big(\{n \Delta t\}_{n=0}^\infty\big)\big] &= \frac{1}{\Delta t} \big[ Z[q\big(\{(n-1) \Delta t\}_{n=0}^\infty\big)] - 2Z[q\big(\{n \Delta t\}_{n=0}^\infty\big)] + Z[q\big(\{(n+1) \Delta t\}_{n=0}^\infty\big)] \big] \\
        &= \frac{1}{\Delta t} \big[ z^{-1} Z[q\big(\{n \Delta t\}_{n=0}^\infty\big)] - 2Z[q\big(\{n \Delta t\}_{n=0}^\infty\big)] + z Z[q\big(\{n \Delta t\}_{n=0}^\infty\big)] \big] \\
        &= \frac{(z-1)^2}{z} \frac{1}{\Delta t} Z[q\big(\{n \Delta t\}_{n=0}^\infty\big)]
    \end{align}
    
    To get $Z[q\big(\{n \Delta t\}_{n=0}^\infty\big)]$, a Z-transform is applied to each sub-element of $q(t)$:
    \begin{align}
        Z\big[q\big(\{n \Delta t\}_{n=0}^\infty\big)\big] &= \frac{K_1 \Delta t z}{(1-z)^2} + \frac{K_2 z}{z-1} + \sum_{k=1}^{\Lambda_R} \frac{\big[\gamma_k e^{r_k \Delta t}\big] z}{\big[e^{r_k \Delta t}\big] z - 1} \\
        & + \sum_{k=1}^{\Lambda_I} \frac{\big[ 2\varsigma_1 e^{2\sigma_1 \Delta t}\big]z^2 + \big[2e^{\sigma_1\Delta t}(\varsigma_2\sin(\sigma_2\Delta t) - \varsigma_1 \cos(\sigma_2 \Delta t))\big] z}{\big[e^{2\sigma_1 \Delta t}\big]z^2 - \big[2e^{\sigma_1 \Delta t}\cos(\sigma_2 \Delta t)\big]z + 1} \\
        &= \frac{K_1 \Delta t z}{(1-z)^2} + \frac{K_2 z}{z-1} + \sum_{k=1}^{\Lambda_R + \Lambda_I} \frac{\mathcal{N}_k}{\mathcal{D}_k} \\
        &= \frac{z}{(1-z)^2}\frac{\mathcal{M}( K_1 \Delta t-K_2 + K_2 z ) + z(1-z^{-1})^2 \overline{\mathcal{M}}}{\mathcal{M}} \label{eq:simp_1}
    \end{align}
    
    \noindent where:
    \begin{align}
        \mathcal{M} &= \prod_{k = 1}^{\Lambda_R + \Lambda_I} \mathcal{D}_k \label{eq:simp_2}\\
        \overline{\mathcal{M}} &= \sum_{k=1}^{\Lambda_R + \Lambda_I} \mathcal{N}_k \prod_{\substack{j = 1 \\ j\neq i}}^{\Lambda_R + \Lambda_I} \mathcal{D}_j \label{eq:simp_3}
    \end{align}
    
    \noindent Now, the coefficients of the z-transfer function are recovered using the following identities:
    \begin{align}
        G_{X}^Z(z) &= \bigg(\frac{\mathcal{M}( K_1 \Delta t-K_2 + K_2 z ) + z(1-z^{-1})^2 \overline{\mathcal{M}}}{\mathcal{M} \ \Delta t}\bigg)_X = \frac{a_0 + a_1 z^{-1} + ... + a_m z^{-m}}{1 + d_1 z^{-1} + .... + d_m z^{-m}} \\
        G_{Y}^Z(z) &= \bigg(\frac{\mathcal{M}( K_1 \Delta t-K_2 + K_2 z ) + z(1-z^{-1})^2 \overline{\mathcal{M}}}{\mathcal{M} \ \Delta t}\bigg)_Y = \frac{b_0 + b_1 z^{-1} + ... + b_m z^{-m}}{1 + d_1 z^{-1} + .... + d_m z^{-m}} \\
        G_{Z}^Z(z) &= \bigg(\frac{\mathcal{M}( K_1 \Delta t-K_2 + K_2 z ) + z(1-z^{-1})^2 \overline{\mathcal{M}}}{\mathcal{M} \ \Delta t}\bigg)_Z = \frac{c_0 + c_1 z^{-1} + ... + c_m z^{-m}}{1 + d_1 z^{-1} + .... + d_m z^{-m}}
    \end{align}

    The response factors can be recovered without explicitly computing the time-domain solutions in (\ref{eq:theta_2}) and (\ref{eq:theta_3}) by computing the Taylor series expansions of the z-transfer functions in terms of $z^{-1}$. To see this, the definition of the z-transfer function is invoked, and we have that the ratio of output (in terms of response factors) to input (in terms of the unit temperature pulses is given by:
    \begin{align}
        G_{\phi}^Z(z) = \frac{Z[\{\phi[k]\}_{k=0}^\infty}{Z[\{\vartheta[k]\}_{k=0}^\infty]} &= \frac{\sum_{k=0}^{\infty} \phi[k]z^{-k}}{1}, \quad \phi \in \{ X,Y,Z  \} \\
        &= \sum_{k=0}^{\infty} \phi[k]z^{-k} \label{eq:yk_z}
    \end{align}

    \noindent where $\vartheta[k]$ is the unit triangular temperature pulse at time $k\Delta t$. Its Z-transform is easily shown to be equivalent to 1:
    \begin{align}
        Z[\{\vartheta[k]\}_{k=0}^\infty] &= \frac{1}{\Delta t} \big[ Z[\widetilde{f}\big(\{(n-1) \Delta t\}_{n=0}^\infty\big)] - 2Z[\widetilde{f}\big(\{n \Delta t\}_{n=0}^\infty\big)] + Z[\widetilde{f}\big(\{(n+1) \Delta t\}_{n=0}^\infty\big)] \big] \\
        &= \frac{(z-1)^2}{z \ \Delta t} Z[\widetilde{f}\big(\{n \Delta t\}_{n=0}^\infty\big)] \\
        &= \frac{(z-1)^2}{z \ \Delta t}\frac{z \ \Delta t}{(z-1)^2} = 1
    \end{align}
    
    \noindent where $\widetilde{f}$ is the unit ramp function. Equating (\ref{eq:yk_z}) to its corresponding CTF z-transfer function, a recurrence relation can be established for the response factors as follows:
    \begin{align}
        &\frac{\sum_{k=0}^m a_{\phi, k} z^{-k}}{\sum_{k=0}^m d_k z^{-k}} = \sum_{k=0}^{\infty} \phi[k]z^{-k}, \quad\quad\ \{a_X\} = \{a\}, \ \{a_Y\} = \{b\}, \ \{a_Z\} = \{c\} \\
        \implies &\begin{cases}
            \phi[k] &= \displaystyle\frac{1}{d_0} \bigg[ a_{\phi,k} - \sum_{j=0}^{k-1} \phi[j] \ d_{k-j} \bigg], \quad\quad\quad k = 0,...,m \\
            \phi[k] &= \displaystyle\frac{1}{d_0} \bigg[ - \sum_{j=0}^{m-1} \phi[k-m+j] \ d_{m-j} \bigg], \quad k = m+1,m+2,...
        \end{cases}
    \end{align}

    \subsection{Implementation Details}
    
    Polynomial additions and multiplications are performed numerically during the simplification processes of (\ref{eq:simp_1}), (\ref{eq:simp_2}) and (\ref{eq:simp_3}) without resorting to complicated analytical expansion formulas such as those based on elementary symmetric polynomials. The linear algebra required to compute the Padé approximants and polynomial roots is performed using the \texttt{LAPACK} routines \texttt{dgesv} and \texttt{zhseqr}. The former solves the linear system via LU-decomposition while the latter exploits the upper Hessenberg structure of the polynomial-root linear system:
    \begin{align}
        \begin{bmatrix}
            -\frac{\alpha_1}{\alpha_0} & -\frac{\alpha_2}{\alpha_0} & -\frac{\alpha_3}{\alpha_0} & ... & -\frac{\alpha_{N-1}}{\alpha_0} & -\frac{\alpha_N}{\alpha_0} \\
            1 & 0 & 0 & ... & 0 & 0 \\
            0 & 1 & 0 & ... & 0 & 0 \\
            \vdots \\
            0 & 0 & 0 & .. & 1 & 0
        \end{bmatrix}
    \end{align}
    
    \noindent to compute its eigenvalues (whose multiplicative inverses are the roots themselves). The coefficients $\alpha_k$ correspond to a polynomial of the form $\sum_{k=0}^N \alpha_k x^k$.

\begin{comment}
\section{Algorithm}

\begin{algorithm}[H]
\SetAlgoLined
\KwIn{Wall/roof data $\{(L_j,\kappa_j,\rho_j,C_{p,j},R_j)\}_{j=0}^{N_{\text{layers}}}$}
\KwResult{Write here the result }
 initialization\;
 \While{While condition}{
  instructions\;
  \eIf{condition}{
   instructions1\;
   instructions2\;
   }{
   instructions3\;
  }
 }
 \caption{Summary of steps to compute conduction transfer function coefficients.}
\end{algorithm}
\end{comment}

\section{Case Studies and Validation}

To validate the CDSE method and its associated implementation, values of CTF coefficients and response factors are compared with those generated by the FDR method and reported in the ASHRAE Handbook (1997) using a variety of test problems. The behavior of the response factors and CTF coefficients can be studied by comparing the frequency characteristics of their associated transfer functions with exact theoretical values by using the $L^2$ error criterion defined by Chen et al. \cite{FDR-Ver II}:
\begin{align}
    E = \frac{1}{U}\sqrt{\frac{1}{N_f} \sum_{i=1}^{N_f} (\psi_i - \bar{\psi_i})^2 }
\end{align}

\noindent where $\psi_i, \bar{\psi_i}$ are the magnitudes of the s-domain external/cross/internal flow transfer functions subjected to a sinusoidal input $s = j \omega_i$ in the discrete, logarithmically-spaced set of frequencies $\Omega = \{ \omega_1, ... , \omega_{N_f} \}$, with $\omega_1 = 10^{-8}$ and $\omega_{N_f}$ being chosen according to the class of wall being considered (a very light wall might require $\omega_{N_f} = 10^{-2}$, while  numerical behavior for a heavyweight wall would become meaningless beyond $\omega_{N_f} = 10^{-4}$). Frequency characteristics of the z-transfer functions are recovered by setting $z^{-1} = e^{-j \omega_i \Delta t}$. This information is also used to generate Bode plots that help assess the region of convergence of the approximated polynomial transfer functions.

An important condition that must be satisfied by the z-transfer functions is that the ratio of the sums of the numerator and denominator coefficients should be equal to the $U$-value:
\begin{align}
    \frac{\sum_{k=0}^m a_k}{\sum_{k=0}^m d_k} = \frac{\sum_{k=0}^m a_k}{\sum_{k=0}^m d_k} = \frac{\sum_{k=0}^m a_k}{\sum_{k=0}^m d_k} = U
\end{align}

This condition is verified in Case Studies I and IV, and can be used in conjunction with the $L^2$ error criterion defined above to form quality control parameters for results generated by the present algorithm which, in terms of tolerances, can be summarized as:
\begin{align}
    \frac{\sum_{k=0}^m a_{\phi,_k}}{\sum_{k=0}^m d_k} &< \epsilon_1, \quad \phi \in \{X, Y, Z\} \\
    E_{\phi} &< \epsilon_2
\end{align}

    \subsection{Case Study I: Brick/Cavity Wall}

    \begin{table}[h!]
        \centering
        \small
        \begin{tabular*}{\textwidth}{l@{\extracolsep{\fill}}ccccc}
            \hline
            Description & \multicolumn{5}{l}{Physical Properties} \\\cline{2-6}
            & $L$ (mm) & $\lambda$ (W m$^{-1}$ K$^{-1}$) & $\rho$ (kg m$^{-3}$ ) & $C_p$ (J kg$^{-1}$ K$^{-1}$) & $R$ (m$^{2}$ K W$^{-1}$) \\
            \hline
            Outside surface film & & & & & 0.060 \\
            Brickwork & 105 & 0.840 & 1700 & 800 & 0.125 \\
            Cavity & & & & & 0.180 \\
            Heavyweight concrete & 100 & 1.630 & 2300 & 1000 & 0.06135 \\
            Inside surface film & & & & & 0.120 \\
            \hline
        \end{tabular*}
        \caption{Physical properties of the brick/cavity wall.}
        \label{tab:brick-cav}
    \end{table}

\begin{figure}[h]
        \centering
        \begin{subfigure}[b]{1\textwidth}
            \includegraphics[scale=1]{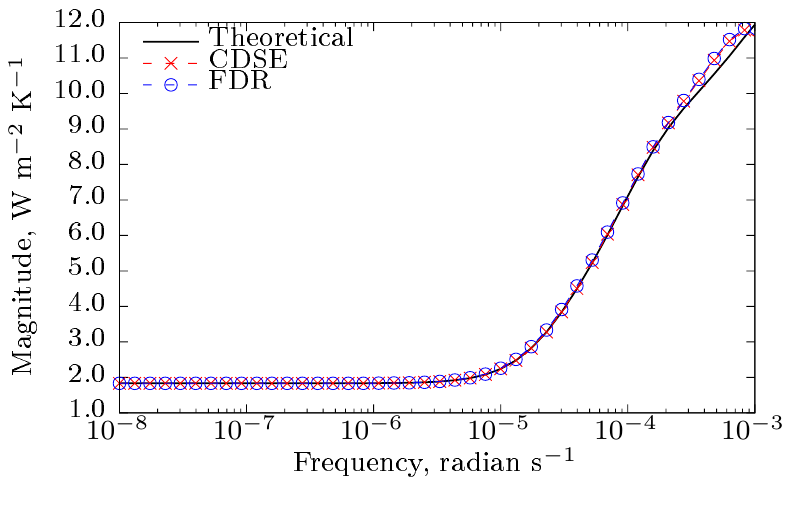}
            \includegraphics[scale=1]{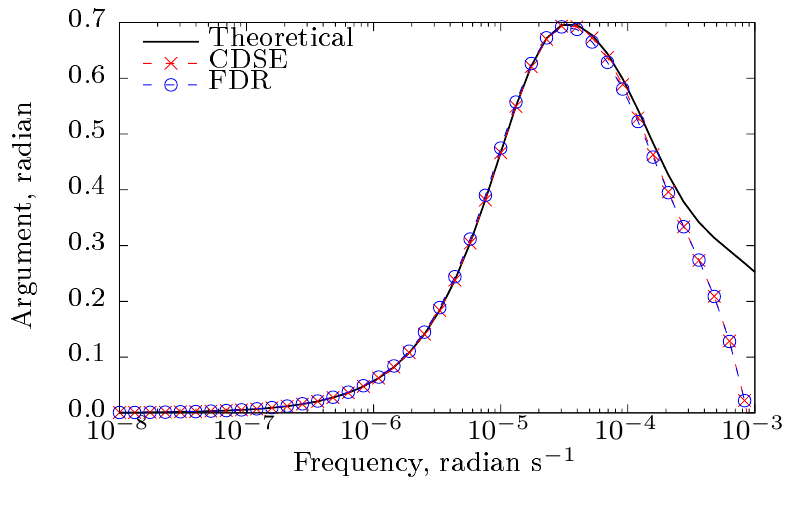}
            \caption{}
            \vspace{0.7cm}
        \end{subfigure}
        \begin{subfigure}[b]{1\textwidth}
            \includegraphics[scale=1]{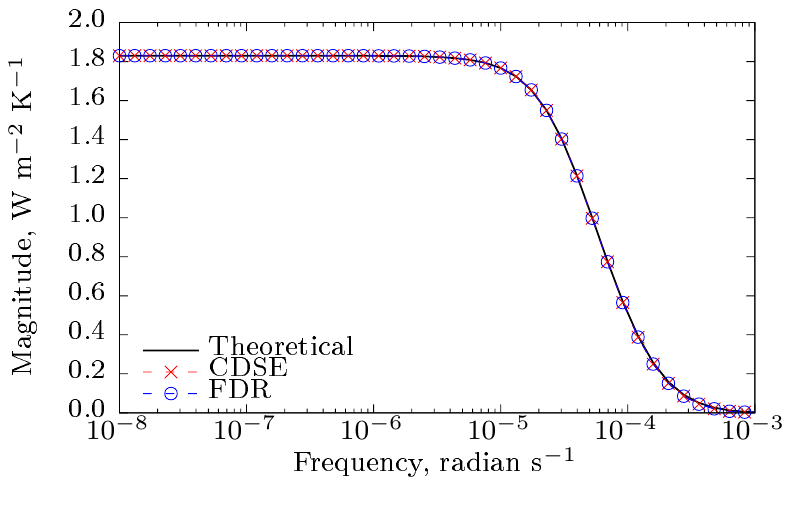}
            \includegraphics[scale=1]{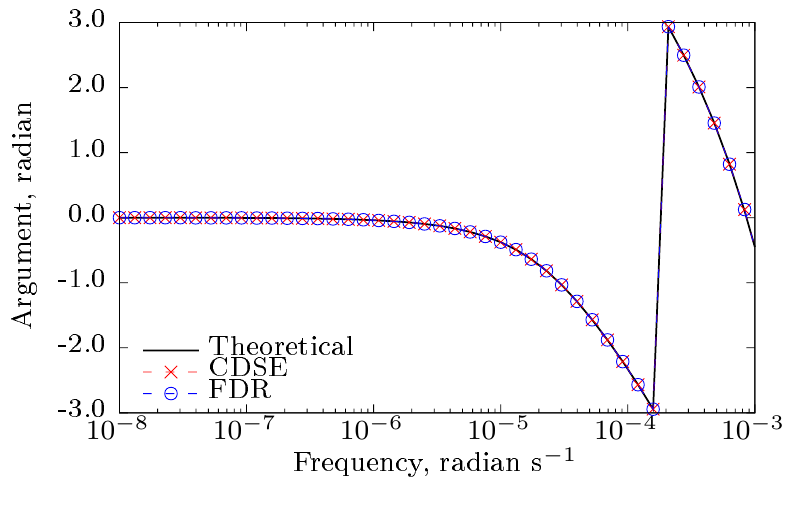}
            \caption{}
            \vspace{0.7cm}
        \end{subfigure}
        \begin{subfigure}[b]{1\textwidth}
            \includegraphics[scale=1]{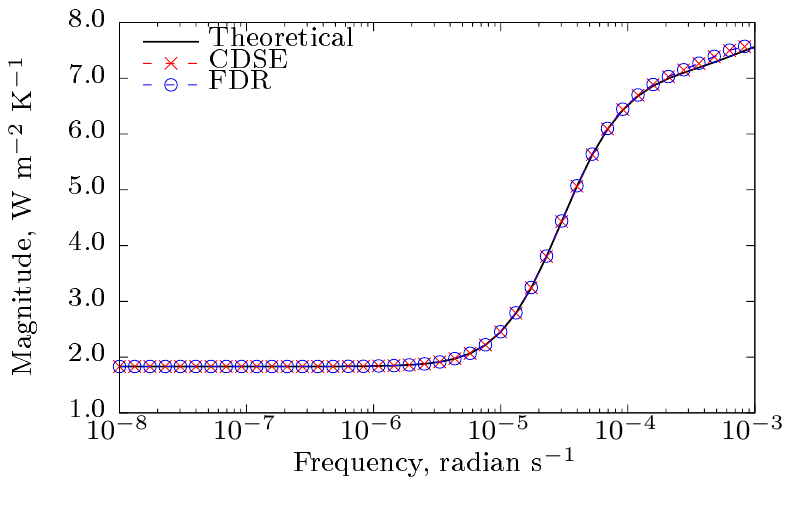}
            \includegraphics[scale=1]{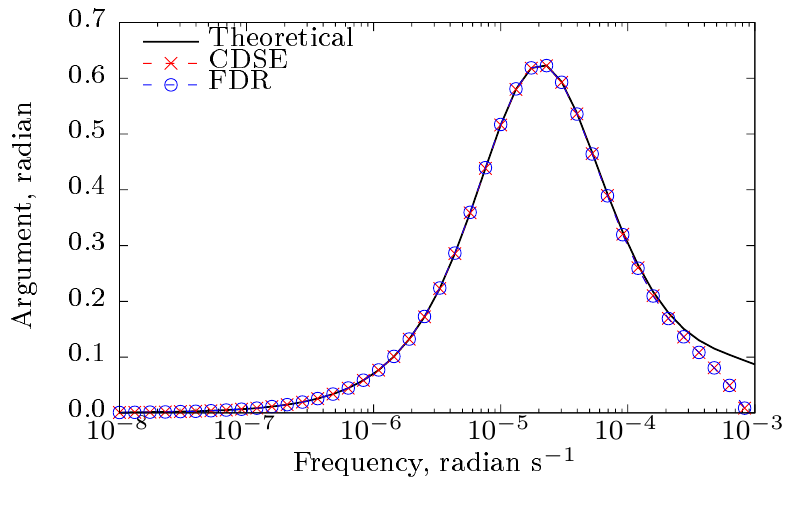}
            \caption{}
            \vspace{0.7cm}
        \end{subfigure}
        \caption{External (a), cross (b) and internal (c) flow Bode diagrams for the brick/cavity wall of Case Study I.}
        \label{fig:cs-i}
    \end{figure}

    The brick/cavity wall test was considered by Xu et al. \cite{FDR Imp.} when validating their improvement of the FDR method with respect to the time domain method of Davies \cite{Davies}. The physical properties of the wall are summarized in Table \ref{tab:brick-cav}. Results of the CDSE method are tabulated and compared with those of the improved FDR method in Table \ref{tab:cs-i}. Bode plots are used to visually assess the frequency characteristics of the z-transfer functions associated with these coefficients with respect to the exact theoretical frequency characteristics over a frequency interval $[10^{-8},10^{-3}]$; these plots are displayed in Figure \ref{fig:cs-i}.
    
    There is a clear agreement between CTF coefficients generated by the CDSE and improved FDR methods, with minor differences on the order of $10^{-2}$ and smaller that can be accounted for by round-off errors in the physical properties of the wall supplied as input. The Bode plots illustrate this strong agreement over the specified frequency interval in regions where the approximated frequency characteristics agree with the theoretical ones and in regions where they begin to diverge from exact values. Agreement in both sets of regions indicates that the polynomial s-transfer functions estimated via Padé approximants are nearly identical to those generated by least squares regression. 

    \begin{table}[h]
        \centering
        \small
        \begin{tabular*}{\textwidth}{c@{\extracolsep{\fill}}llllllc}
            \hline
            k & 0 & 1 & 2 & 3 & 4 & 5 & $\sum$ \\
            \hline
            $a_k^a$ & 9.547772 & -18.534215 & 10.584111 & -1.585679 & 0.065340 & -0.000761 & 0.076568 \\
            $a_k^b$ & 9.589008 & -18.586934 & 10.586818 & -1.560687 & 0.049680 & -0.001181 & 0.076703 \\
            \\
            $b_k^a$ & 0.000179 & 0.013914 & 0.043449 & 0.018001 & 0.001021 & 0.000005 & 0.076568 \\
            $b_k^b$ & 0.000178 & 0.013914 & 0.043475 & 0.018078 & 0.001052 & 0.000006 & 0.076703 \\
            \\
            $c_k^a$ & 6.953532 & -12.228572 & 5.995942 & -0.665431 & 0.021226 & -0.000128 & 0.076568 \\
            $c_k^b$ & 6.959635 & -12.226366 & 5.979019 & -0.653456 & 0.018078 & -0.000207 & 0.076703 \\
            \\
            $d_k^a$ & 1.000000 & -1.621649 & 0.727483 & -0.065668 & 0.001670 & -0.000004 & 0.041833 \\
            $d_k^b$ & 1.000000 & -1.619844 & 0.724520 & -0.064306 & 0.001543 & -0.000006 & 0.041906 \\
            \hline
        \end{tabular*}
        \caption{Comparison of CTF coefficients for the brick/cavity wall generated by the CDSE and FDR methods.}
        \label{tab:cs-i}
        $^a$ Complex domain series expansion method. \\
        $^b$ Frequency-domain regression method.
    \end{table}

    \subsection{Case Study II: Comparison with ASHRAE Handbook (1997)}
    
    The 1997 ASHRAE Handbook \cite{ASHRAE} provides a tabulation of CTF coefficients for 42 roof and 41 wall constructions, converted to SI from Harris and McQuiston's original study on the categorization of roofs and walls based on thermal response \cite{McQuiston}. For each multilayer element, 7 coefficients for the cross flow z-transfer function numerator and denominator are given. These conductive elements serve as an excellent basis of comparison, as they cover a wide weight range.
    
    The results of a test run over all roofs and walls are tabulated in Table \ref{tab:cs-ii}. The $L^2$ error criterion is invoked to compare the frequency characteristics of the cross flow z-transfer function reported by the Handbook and generated by the present method with exact theoretical characteristics. A frequency interval of $[10^{-8},10^{-3}]$ is specified and discretized with $N_f = 100$ for the purposes of computing these frequency characteristics, and $m = 6$ is fixed throughout the test run.
    
    It is clear that the CDSE method is capable of computing accurate z-transfer functions over the full range of wall weights, with the largest $L^2$ error among both roofs and walls being 3.43\%. The error is also seen to decrease as wall weight increases (heavier roofs and walls are generally found nearer to the end of the Handbook tables, however, the order is not strict). For lighter roofs and walls, the $L^2$ errors for ASHRAE's coefficients and those generated by the present algorithm are seen to be nearly equal. Beyond the first few roofs and walls, the errors for the CDSE method remain below 1\% consistently. For heavier walls, ASHRAE's z-transfer functions begin to fail, with notably high errors being associated with Walls 4, 34, 37 and 38 (whose errors are 57.73\%, 64.19\%, 21.99\% and 24.24\%, respectively). The present solver's capability of handling all given roofs and walls demonstrates robustness with respect to wall weight.
    
    A test run with $m = 10$ verified that the linear systems remain well-conditioned, as it was found that the $L^2$ errors were nearly the same as those computed with $m = 6$. This also indicates that a default choice of $m = 6$ will likely be able to hand roofs and walls encountered in practice.
    
    \begin{table}[h]
        \centering
        \small
        \begin{tabular*}{\textwidth}{ccccccccccccc}
            \hline
            \multicolumn{6}{c}{Roofs} & \hspace{0.5cm} & \multicolumn{6}{c}{Walls} \\ \cline{1-6} \cline{8-13}
            No. & ASHRAE & CDSE & No. & ASHRAE & CDSE & & No. & ASHRAE & CDSE & No. & ASHRAE & CDSE \\
            \hline
1 & 2.87\% & 2.87\% & 22 & 0.07\% & 0.07\% & &1 & 3.43\% & 3.43\% & 22 & 0.14\% & 0.09\% \\
2 & 2.24\% & 2.24\% & 23 & 3.96\% & 0.06\% & &2 & 1.22\% & 1.22\% & 23 & 0.66\% & 0.07\% \\
3 & 0.98\% & 0.98\% & 24 & 0.12\% & 0.05\% & &3 & 1.05\% & 1.05\% & 24 & 0.13\% & 0.06\% \\
4 & 0.59\% & 0.59\% & 25 & 0.23\% & 0.04\% & &4 & 57.73\% & 1.19\% & 25 & 0.64\% & 0.05\% \\
5 & 0.57\% & 0.57\% & 26 & 0.20\% & 0.09\% & &5 & 0.49\% & 0.49\% & 26 & 0.10\% & 0.04\% \\
6 & 0.60\% & 0.60\% & 27 & 0.06\% & 0.06\% & &6 & 0.43\% & 0.43\% & 27 & 0.65\% & 0.03\% \\
7 & 0.49\% & 0.49\% & 28 & 0.20\% & 0.05\% & &7 & 0.40\% & 0.40\% & 28 & 0.35\% & 0.03\% \\
8 & 0.41\% & 0.41\% & 29 & 0.14\% & 0.04\% & &8 & 0.25\% & 0.25\% & 29 & 1.06\% & 0.02\% \\
9 & 0.35\% & 0.34\% & 30 & 1.20\% & 0.04\% & &9 & 0.26\% & 0.26\% & 30 & 2.62\% & 0.05\% \\
10 & 0.23\% & 0.23\% & 31 & 0.22\% & 0.03\% & &10 & 0.21\% & 0.21\% & 31 & 4.59\% & 0.05\% \\
11 & 0.19\% & 0.19\% & 32 & 0.70\% & 0.03\% & &11 & 0.18\% & 0.18\% & 32 & 2.44\% & 0.04\% \\
12 & 0.18\% & 0.18\% & 33 & 1.31\% & 0.02\% & &12 & 0.17\% & 0.16\% & 33 & 0.51\% & 0.03\% \\
13 & 0.19\% & 0.19\% & 34 & 0.36\% & 0.02\% & &13 & 0.14\% & 0.14\% & 34 & 64.19\% & 0.02\% \\
14 & 0.15\% & 0.15\% & 35 & 1.43\% & 0.04\% & &14 & 0.17\% & 0.10\% & 35 & 12.30\% & 0.27\% \\
15 & 0.15\% & 0.10\% & 36 & 0.60\% & 0.03\% & &15 & 0.23\% & 0.15\% & 36 & 0.75\% & 0.02\% \\
16 & 0.08\% & 0.08\% & 37 & 0.69\% & 0.03\% & &16 & 0.12\% & 0.11\% & 37 & 21.99\% & 0.03\% \\
17 & 0.20\% & 0.19\% & 38 & 4.70\% & 0.02\% & &17 & 0.85\% & 0.10\% & 38 & 24.24\% & 0.03\% \\
18 & 0.37\% & 0.11\% & 39 & 1.16\% & 0.02\% & &18 & 0.28\% & 0.08\% & 39 & 9.73\% & 0.02\% \\
19 & 0.13\% & 0.13\% & 40 & 0.39\% & 0.02\% & &19 & 0.25\% & 0.07\% & 40 & 10.71\% & 0.02\% \\
20 & 0.12\% & 0.11\% & 41 & 3.09\% & 0.01\% & &20 & 0.37\% & 0.06\% & 41 & 8.83\% & 0.01\% \\
21 & 0.28\% & 0.09\% & 42 & 6.58\% & 0.01\% & &21 & 0.15\% & 0.04\% & - & - & - \\
            \hline
        \end{tabular*}
        \caption{A comparison of the $L^2$ error norms associated with the CTF coefficients reported by ASHRAE and generated by the CDSE method (with $m = 6$) for a set of 42 roof and 41 wall constructions.}
        \label{tab:cs-ii}
    \end{table}
    
    \subsection{Case Study III: A Heavyweight Wall}
    
\begin{figure}[h]
        \centering
        \includegraphics[scale=1]{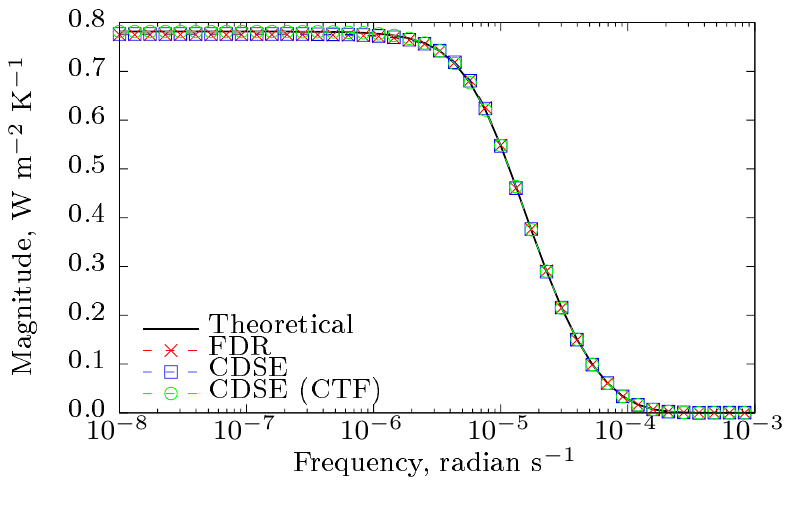}
        \includegraphics[scale=1]{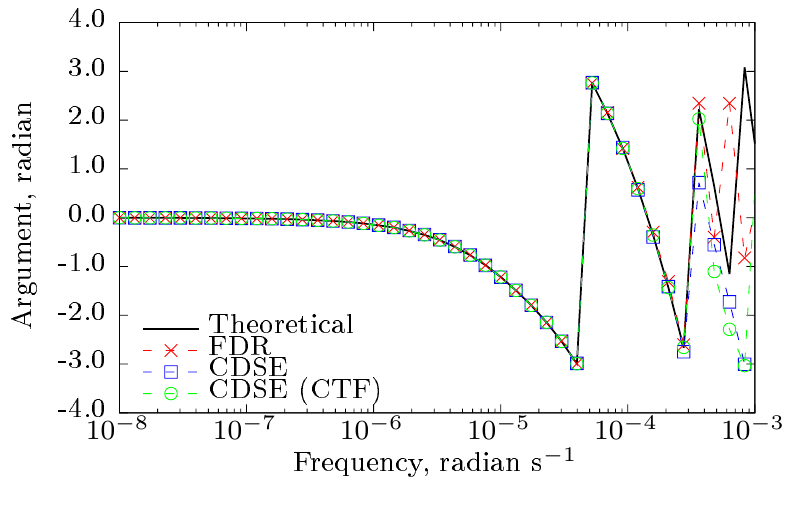}
        \caption{Cross flow Bode diagrams for the heavyweight wall of Case Study III.}
        \label{fig:cs-iii-plots}
    \end{figure}

    The physical properties of a heavyweight wall reported to be commonly used in China \cite{FDR-Ver II} are displayed in Table \ref{tab:cs-iii}. The response factors for cross heat flow computed by the CDSE method with $m = 6$ are compared with the set generated by the FDR method and reported by Chen et al.  by visual inspection of the coefficient values (the first 72 of which are listed in Table \ref{tab:cs-iii-yk}) and the Bode plots of their corresponding z-transfer functions over a frequency range $[10^{-8}, 10^{-3}]$ displayed in Figure \ref{fig:cs-iii-plots} (curves of the CTF z-transfer function are included for reference). 144-term response factor sequences are used to generate the Bode plots. Additionally, the $L^2$ error for the z-transfer functions constructed by both sets of factors are tabulated in Table \ref{tab:cs-iii-err} under varying truncation limits. The error terms are evaluated using an interval $[10^{-9}, 10^{-3}]$ and $N_f = 50$.
    
    \begin{table}[h]
        \centering
        \begin{tabular}{cll}
            \hline
            $k$ & $E_k^a$ & $E_k^b$ \\
            \hline
            72 & 9.125\% & 7.065\% \\
            96 & 3.740\% & 2.905\% \\
            120 & 1.535 & 1.195\% \\
            144 & 0.632\% & 0.492\% \\
            \hline
        \end{tabular}
        \caption{Comparison of $L^2$ errors for response factor z-transfer functions of varying truncation limits for the CDSE and FDR methods.}
        \label{tab:cs-iii-err}
        \small
        $^a$ Complex domain series expansion method. \\
        $^b$ Frequency-domain regression method.
    \end{table}
    
    The z-transfer function curves are seen to agree well with the theoretical frequency characteristics over $[10^{-8},10^{-3.2}]$. Beyond this region, all phase curves begin to diverge from the theoretical curve (on the other hand, magnitude curves continue to vanish). The CDSE and FDR response factor sequences are very similar (with errors on the order of 0.001 due to round-off errors in the wall parameters) except for the first four terms, the variance of which is due to sensitivity in the computation of the CTF z-transfer function. Since the first term obtained from the recurrence relation between the CTF and response factor z-transfer functions will always be $Y[0] = b_0 / d_0 = b_0$, round-off errors in the wall parameters will be more apparent in the initial few sequence terms. The $L^2$ error terms reveal that the variation in the coefficient values does not impact accuracy of the overall sequence, with errors for the CDSE method turning out to be slightly smaller than those of the FDR method (which, for confirmation, were found to be nearly identical to those reported in \cite{FDR-Ver II}).
    
    \begin{table}[h]
        \centering
        \small
        \begin{tabular*}{\textwidth}{l@{\extracolsep{\fill}}ccccc}
            \hline
            Description & \multicolumn{5}{l}{Physical Properties} \\\cline{2-6}
            & $L$ (mm) & $\lambda$ (W m$^{-1}$ K$^{-1}$) & $\rho$ (kg m$^{-3}$ ) & $C_p$ (J kg$^{-1}$ K$^{-1}$) & $R$ (m$^{2}$ K W$^{-1}$) \\
            \hline
            Outside surface film & & & & & 0.0538 \\
            Common brick & 370 & 0.814 & 1800 & 879 &  \\
            Foam concrete & 100 & 0.209 & 600 & 837 &  \\
            Wood wool board & 25 & 0.163 & 400 & 2093 &  \\
            Stucco & 20 & 0.814 & 1600 & 837 & \\
            Inside surface film & & & & & 0.1147 \\
            \hline
        \end{tabular*}
        \caption{Physical properties of the heavyweight wall.}
        \label{tab:cs-iii}
    \end{table}
    
    \begin{table}[h]
        \centering
        \footnotesize
        \begin{tabular*}{\textwidth}{c@{\extracolsep{\fill}}llcll}
            \hline
            k & $Y_k^a$ & $Y_k^b$ & k & $Y_k^a$ & $Y_k^b$  \\
            \hline
            0 & -0.0000263957 & 0.0000196345 & 36 & 0.0121633014 & 0.0121727560 \\
            1 & 0.0000675681 & 0.0000108668 & 37 & 0.0117379521 & 0.0117479420 \\
            2 & -0.0000468852 & 0.0000040855 & 38 & 0.0113253164 & 0.0113356370 \\
            3 & 0.0000004003 & 0.0000231081 & 39 & 0.0109254660 & 0.0109359310 \\
            4 & 0.0001465373 & 0.0001405920 & 40 & 0.0105383625 & 0.0105488070 \\
            5 & 0.0006073107 & 0.0006138580 & 41 & 0.0101638835 & 0.0101741660 \\
            6 & 0.0016778107 & 0.0016942230 & 42 & 0.0098018425 & 0.0098118420 \\
            7 & 0.0033991044 & 0.0034144010 & 43 & 0.0094520055 & 0.0094616240 \\
            8 & 0.0055920679 & 0.0056038820 & 44 & 0.0091141032 & 0.0091232590 \\
            9 & 0.0079971773 & 0.0080073320 & 45 & 0.0087878415 & 0.0087964740 \\
            10 & 0.0103774754 & 0.0103869760 & 46 & 0.0084729095 & 0.0084809710 \\
            11 & 0.0125625475 & 0.0125711980 & 47 & 0.0081689851 & 0.0081764420 \\
            12 & 0.0144529693 & 0.0144606400 & 48 & 0.0078757406 & 0.0078825730 \\
            13 & 0.0160069035 & 0.0160140340 & 49 & 0.0075928461 & 0.0075990410 \\
            14 & 0.0172224324 & 0.0172296410 & 50 & 0.0073199722 & 0.0073255280 \\
            15 & 0.0181219045 & 0.0181294380 & 51 & 0.0070567927 & 0.0070617130 \\
            16 & 0.0187402231 & 0.0187477510 & 52 & 0.0068029861 & 0.0068072800 \\
            17 & 0.0191169751 & 0.0191237540 & 53 & 0.0065582369 & 0.0065619190 \\
            18 & 0.0192915961 & 0.0192967990 & 54 & 0.0063222362 & 0.0063253250 \\
            19 & 0.0193006888 & 0.0193036980 & 55 & 0.0060946828 & 0.0060971980 \\
            20 & 0.0191767574 & 0.0191773220 & 56 & 0.0058752838 & 0.0058772480 \\
            21 & 0.0189478135 & 0.0189460690 & 57 & 0.0056637540 & 0.0056651920 \\
            22 & 0.0186374833 & 0.0186338750 & 58 & 0.0054598172 & 0.0054607550 \\
            23 & 0.0182653747 & 0.0182605400 & 59 & 0.0052632054 & 0.0052636670 \\
            24 & 0.0178475567 & 0.0178422040 & 60 & 0.0050736593 & 0.0050736720 \\
            25 & 0.0173970623 & 0.0173918750 & 61 & 0.0048909278 & 0.0048905160 \\
            26 & 0.0169243691 & 0.0169199390 & 62 & 0.0047147683 & 0.0047139590 \\
            27 & 0.0164378325 & 0.0164346180 & 63 & 0.0045449464 & 0.0045437640 \\
            28 & 0.0159440641 & 0.0159423810 & 64 & 0.0043812354 & 0.0043797050 \\
            29 & 0.0154482543 & 0.0154482800 & 65 & 0.0042234167 & 0.0042215610 \\
            30 & 0.0149544418 & 0.0149562290 & 66 & 0.0040712792 & 0.0040691220 \\
            31 & 0.0144657381 & 0.0144692410 & 67 & 0.0039246191 & 0.0039221830 \\
            32 & 0.0139845107 & 0.0139896060 & 68 & 0.0037832398 & 0.0037805450 \\
            33 & 0.0135125326 & 0.0135190440 & 69 & 0.0036469516 & 0.0036440200 \\
            34 & 0.0130511042 & 0.0130588220 & 70 & 0.0035155716 & 0.0035124220 \\
            35 & 0.0126011508 & 0.0126098510 & 71 & 0.0033889233 & 0.0033855740 \\
            \hline
        \end{tabular*}
        \caption{The first 71 response factors for the heavyweight wall generated by the CDSE and FDR methods.}
        \label{tab:cs-iii-yk}
        $^a$ Complex domain series expansion method. \\
        $^b$ Frequency-domain regression method.
    \end{table}
    
    \begin{table}[h]
        \centering
        \small
        \begin{tabular*}{\textwidth}{l@{\extracolsep{\fill}}ccccc}
            \hline
            Description & \multicolumn{5}{l}{Physical Properties} \\\cline{2-6}
            & $L$ (mm) & $\lambda$ (W m$^{-1}$ K$^{-1}$) & $\rho$ (kg m$^{-3}$ ) & $C_p$ (J kg$^{-1}$ K$^{-1}$) & $R$ (m$^{2}$ K W$^{-1}$) \\
            \hline
            Outside surface film & & & & & 0.060 \\
            Stucco & 25 & 0.692 & 1858 & 840 & 0.036 \\
            Insulation & 125 & 0.043 & 91 & 840 & 2.907 \\
            Plaster or gypsum & 20 & 0.727 & 1602 & 840 & 0.028 \\
            Inside surface film & & & & & 0.120 \\
            \hline
        \end{tabular*}
        \caption{Physical properties of ASHRAE Wall Group 2.}
        \label{tab:cs-iv}
    \end{table}
    
    \subsection{Case Study IV: Varying the Time Step}

\begin{figure}[h]
        \centering
        \begin{subfigure}[b]{1\textwidth}
            \includegraphics[scale=1]{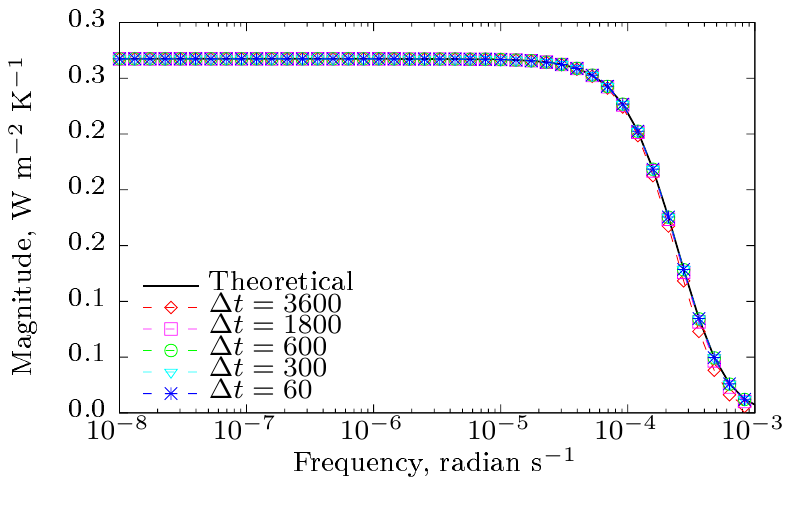}
            \includegraphics[scale=1]{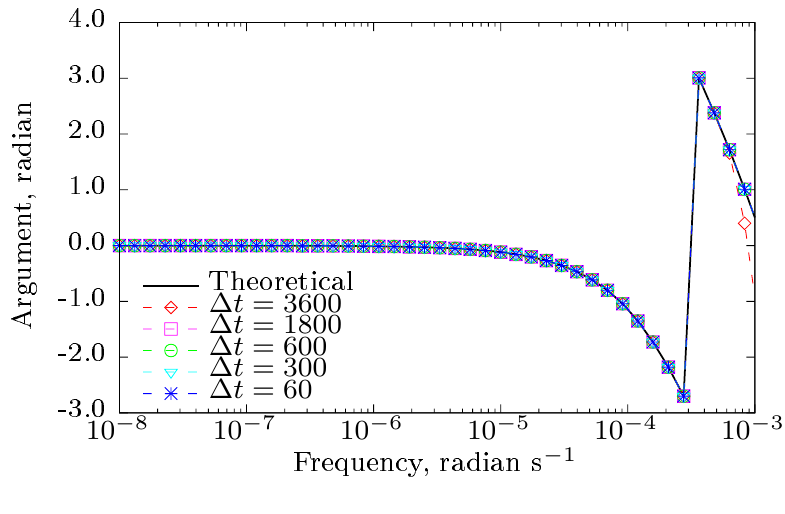}
            \caption{}
            \vspace{0.7cm}
        \end{subfigure}
        \caption{Cross flow Bode diagrams for Wall Group 2 of Case Study IV.}
        \label{fig:cs-iv-plots}
    \end{figure}

    Although hourly analysis (i.e. with a time step of $\Delta t = 3600$ s) is standard for heat load calculations in the building services, other contexts may involve analyses in which smaller time steps are required (such as control simulations of air conditioning systems). Wang et al. \cite{FDR II} investigated the performance of the FDR method as the time step was shortened, noting that previously established methods such as Direct Root Finding and State Space required a larger number of coefficients to sufficiently simulate heat conduction when $\Delta t$ was made small enough, leading to unstable behaviour in the resulting z-transfer functions. CTF coefficients for  Wall Group 2 of the ASHRAE Handbook (whose physical properties are summarized in Table \ref{tab:cs-iv}) were reported for time steps $\Delta t = 3600, 1800, 1200, 900, 600, 300$ and $60$ s. CTF coefficients computed using the CDSE method with a choice of $m = 5$ are tabulated along side those generated by the FDR method in Table \ref{tab:cs-iv-ctf}, with results for time steps $\Delta t = 1200$ and $900$ s  omitted for brevity. The frequency characteristics of corresponding z-transfer functions are represented by Bode plots in Figure \ref{fig:cs-iv-plots} in comparison with theoretical values.
    
    A visual inspection of the Bode plots reveals that the CDSE method remains accurate when the time step is reduced. In fact, the frequency characteristics of the CTF z-transfer functions are more inline with theoretical values when the time step is smaller. The $L^2$ errors for time steps $3600, 1800, 600, 300$ and $60$ s were found to be 1.174\%, 0.305\%, 0.0345\%, 8.622e-3\% and 3.457e-4\%, respectively. For time steps smaller than $60$ s, the error begins to increase rapidly. For example, the error for $\Delta t = 10 s$ was found to be 1.819\%. There is a general agreement between the CTF coefficients computed using the FDR and CDSE methods, with the most significant variations appearing strictly in the $b_k$ coefficients. $U$-values associated with both sets of coefficients are seen to match the true $U$-value of the Wall Group.

    \begin{table}[h]
        \centering
        \scriptsize
        \begin{tabular*}{\textwidth}{c@{\extracolsep{\fill}}llllllc}
            \hline
            k & 0 & 1 & 2 & 3 & 4 & 5 & $U$ \\
            \hline
            \multicolumn{8}{l}{Time step $\Delta t = 3600$ s} \\
            $b_k^a$ & 9.238270E-04 & 3.134899E-02 & 5.424187E-02 & 1.188745E-02 & 2.738740E-04 & 2.703650E-07 & 0.317398 \\
            $b_k^b$ & 9.356678E-04 & 3.165971E-02 & 5.460031E-02 & 1.193587E-02 & 2.768229E-04 & 3.041151E-07 & 0.317479 \\
            $c_k^a$ & 4.964126E+00 & -7.538352E+00 & 3.020772E+00 & -3.499972E-01 & 2.131200E-03 & -3.968320E-06 & 0.317398 \\
            $c_k^b$ & 4.950058E+00 & -7.483352E+00 & 2.971137E+00 & -3.401356E-01 & 1.706114E-03 & -5.092724E-06 & 0.317479 \\
            $d_k^a$ & 1.000000E+00 & -9.408329E-01 & 2.774543E-01 & -2.585314E-02 & 1.226296E-04 & -2.377552E-08 * - \\
            $d_k^b$ & 1.000000E+00 & -9.355651E-01 & 2.738394E-01 & -2.526493E-02 & 1.095122E-04 & -3.532504E-08 & - \\
            \multicolumn{8}{l}{Time step $\Delta t = 1800$ s} \\
            $b_k^a$ & 3.523998E-06 & 1.834509E-03 & 1.144690E-02 & 9.920927E-03 & 1.424604E-03 & 2.253814E-05 & 0.317398 \\
            $b_k^b$ & 4.727864E-06 & 1.853722E-03 & 1.158368E-02 & 1.000414E-02 & 1.438610E-03 & 2.348765E-05 & 0.317479 \\
            $c_k^a$ & 6.104093E+00 & -1.284061E+01 & 8.882510E+00 & -2.266845E+00 & 1.476614E-01 & -2.155427E-03 & 0.317398 \\
            $c_k^b$ & 6.080903E+00 & -1.275364E+01 & 8.788538E+00 & -2.234344E+00 & 1.460655E-01 & -2.616704E-03 & 0.317479 \\
            $d_k^a$ & 1.000000E+00 & -1.730097E+00 & 1.026203E+00 & -2.322156E-01 & 1.393706E-02 & -1.541931E-04 & - \\
            $d_k^b$ & 1.000000E+00 & -1.725393E+00 & 1.020708E+00 & -2.306784E-01 & 1.400800E-02 & -1.879496E-04 & - \\
            \multicolumn{8}{l}{Time step $\Delta t = 600$ s} \\
            $b_k^a$ & 1.888654E-05 & -9.879201E-05 & 2.288415E-04 & -3.73172E-05 & 4.788933E-04 & 1.967507E-04 & 0.317398 \\
            $b_k^b$ & 1.445008E-05 & -7.594148E-05 & 1.830175E-04 & 7.711732E-06 & 4.688070E-04 & 1.867713E-04 & 0.317536 \\
            $c_k^a$ & 7.154002E+00 & -2.330286E+01 & 2.899612E+01 & -1.704157E+01 & 4.666951E+00 & -4.718573E-01 & 0.317398 \\
            $c_k^b$ & 7.120910E+00 & -2.327280E+01 & 2.911951E+01 & -1.727687E+01 & 4.812082E+00 & -5.020521E-01 & 0.315547 \\
            $d_k^a$ & 1.000000E+00 & -3.099307E+00 & 3.687662E+00 & -2.082222E+00 & 5.499704E-01 & -5.362348E-02 & - \\
            $d_k^b$ & 1.000000E+00 & -3.110953E+00 & 3.723589E+00 & -2.123080E+00 & 5.701970E-01 & -5.728142E-02 & - \\
            \multicolumn{8}{l}{Time step $\Delta t = 300$ s} \\
            $b_k^a$ & -8.604667E-06 & 8.803074E-05 & -3.226204E-04 & 5.804443E-04 & -5.418111E-04 & 2.512651E-04 & 0.317398 \\
            $b_k^b$ & -1.127836E-05 & 9.493795E-05 & -3.165406E-04 & 5.426042E-04 & -4.961109E-04 & 2.324727E-04 & 0.317479 \\
            $c_k^a$ & 7.479679E+00 & -2.937890E+01 & 4.551999E+01 & -3.472326E+01 & 1.301909E+01 & -1.916569E+00 & 0.317398 \\
            $c_k^b$ & 7.444271E+00 & -2.934634E+01 & 4.568333E+01 & -3.506161E+01 & 1.325239E+01 & -1.972001E+00 & 0.317479 \\
            $d_k^a$ & 1.000000E+00 & -3.840745E+00 & 5.826007E+00 & -4.355829E+00 & 1.602282E+00 & -2.315674E-01 & - \\
            $d_k^b$ & 1.000000E+00 & -3.855263E+00 & 5.876049E+00 & -4.420216E+00 & 1.638910E+00 & -2.393354E-01 & - \\
            \multicolumn{8}{l}{Time step $\Delta t = 60$ s} \\
            $b_k^a$ & -1.550119E-04 & 8.435387E-04 & -1.841489E-03 & 2.016482E-03 & -1.107979E-03 & 2.444852E-04 & 0.317398 \\
            $b_k^b$ & -1.426411E-04 & 7.742094E-04 & -1.686415E-03 & 1.843344E-03 & -1.011445E-03 & 2.229736E-04 & 0.317478 \\
            $c_k^a$ & 7.769836E+00 & -3.683664E+01 & 6.981223E+01 & -6.611053E+01 & 3.128182E+01 & -5.916703E+00 & 0.317369 \\
            $c_k^b$ & 7.733101E+00 & -3.670617E+01 & 6.965236E+01 & -6.604659E+01 & 3.129529E+01 & -5.927998E+00 & 0.317486 \\
            $d_k^a$ & 1.000000E+00 & -4.721551E+00 & 8.911928E+00 & -8.405527E+00 & 3.961488E+00 & -0.746338E-01 & - \\
            $d_k^b$ & 1.000000E+00 & -4.727177E+00 & 8.933721E+00 & -8.437176E+00 & 3.981911E+00 & -7.512792E-01 & - \\
            \hline
        \end{tabular*}
        \caption{Comparison of CTF coefficients for Wall Group 2 generated by the CDSE and FDR method with various choices of time step $\Delta t$.}
        \label{tab:cs-iv-ctf}
        $^a$ Complex domain series expansion method. \\
        $^b$ Frequency-domain regression method.
    \end{table}

\section{Conclusion}

An algorithm has been presented for computing response factors and CTF coefficients based on series expansions of the solutions of the governing equations in the $s$-domain. This approach was motivated by the recently developed frequency-domain regression approach in that external, cross and internal heat flow transfer functions are approximated with rational functions that simplify the Laplace inversion process. However, the rational functions are estimated using Padé approximants rather than solving a weighted least squares problem. The method is mathematically straightforward, and its open source C++ implementation only requires two \texttt{FORTRAN} routines to perform the necessary linear algebra.

A set of case studies were chosen to study the performance and results of the CDSE method, which in many ways can be directly compared with those of the FDR method due to the fact that the two methods generate 'optimal' polynomial approximations of the systems' transfer functions. The behaviour of the z-transfer functions was found to be nearly identical in both methods, regardless of the choice of time step $\Delta t$, however, some differences in the coefficient values (especially in the $b_k$) was detected. Nonetheless, comparisons of z-transfer function frequency characteristics with theoretical values and with characteristics of other results reported in ASHRAE's Handbook and in the literature demonstrate a high degree of accuracy and robustness with respect to the potential range of wall weights encountered in practice. 

\section{Funding}

This research did not receive any specific grant from funding agencies in the public, commercial, or not-for-profit sectors.

\section{Acknowledgements}

I would like to thank Kassem Jaber for assisting me with technical details regarding the dynamic system analysis and in assessing the overall composition of this paper.

%\appendix

% \bibliographystyle{elsarticle-num} 
% \bibliography{cas-refs}


\end{document}